\documentclass[12pt]{iopart}

\usepackage{iopams}
\usepackage{graphicx}
\usepackage{subfig}
\usepackage{color}
\usepackage{ulem}

\begin{document}

\title{Quasi-linear theory of forced magnetic reconnection for the transition from linear to Rutherford regime}

\author{Wenlong HUANG}
\address{School of Computer Science and Technology,\\
Anhui  Engineering Laboratory for  Industrial Internet  Intelligent Applications and Security, Anhui University of Technology, Ma'anshan, Anhui 243002, China}

\author{Ping ZHU}
\address{International Joint Research Laboratory of Magnetic Confinement Fusion and Plasma Physics, State Key Laboratory of Advanced Electromagnetic Engineering and Technology, School of Electrical and Electronic Engineering, Huazhong University of Science and Technology, Wuhan, Hubei 430074, China}
\address{Department of Engineering Physics, University of Wisconsin-Madison, Madison, Wisconsin 53706, USA}
\ead{zhup@hust.edu.cn}

\newpage

\begin{abstract}
Using the in-viscid two-field reduced MHD model, a new analytical theory is developed to unify the Hahm-Kulsrud-Taylor (HKT) linear solution and the Rutherford quasi-linear regime. Adopting a quasi-linear approach, we obtain a closed system of equations for plasma response in a static plasma in slab geometry. An integral form of analytical solution is obtained for the forced magnetic reconnection, uniformly valid throughout the entire regimes from the HKT linear solution to the Rutherford quasi-linear solution. In particular, the quasi-linear effect can be described by a single coefficient $K_s\propto S^{8/5} \psi_c^2$, where $S=\frac{\tau_R}{\tau_A}$ and $\psi_c$ are the Lunquist number and amplitude of external magnetic perturbation, respectively. The HKT linear solution for response can be recovered when the index $K_s\rightarrow 0$. On the other hand, the quasi-linear effect plays a key role in the island growth when $K_s\sim1$. Our new analytical solution has also been compared with reduced MHD simulations with agreement.
\end{abstract}

\maketitle

\section{Introduction}
In the last decades, three dimensional external magnetic coils have been widely equipped in various fusion devices due to the emerging and promising potential for controlling plasma activities~\cite{hu12a, hu13a, chen18a, ding2018}. For example, various MHD instabilities, such as edge localized modes (ELMs), can be suppressed or mitigated by resonant magnetic perturbation (RMP) coils~\cite{evans05a, sut11a, jak09, sun16, yang2016, yang2019a}. Recent experimental and theoretical results show that the mechanism of ELM suppression or mitigation by RMPs is connected with the forced magnetic reconnection (FMR) processes on resonant surfaces in the pedestal region~\cite{hu19, fitz2019, fitz2020, fitz2020c}.

According to the island width, which is decided by the amplitude of RMP, there are several different models of FMR. When the island width ($W$) is much narrower than the resistive layer width ($\delta_{\rm layer}$), quasi-linear magnetic terms can be neglected. In this regime, Hahm and Kulsrud proposed the linear FMR solution in the Taylor problem (HKT)~\cite{hahm85}. Later, the linear HKT solution in the resistive-inertial (RI) regime was extended to include the steady state plasma flow~\cite{fitz91a}. It should be noted that the plasma flow can also be dramatically modified by RMPs even in the regime where $W \ll \delta_{\rm layer}$~\cite{fitz98a, beidler17a, beidler18a}. To study the plasma response in the small island regime, a quasi-linear model for plasma flow response to RMPs in a tokamak has been self-consistently developed purely from the two-field reduced MHD model~\cite{huang20a, huang20b}.

When $W \sim \delta_{\rm layer}$, the forced island enters into the Rutherford regime~\cite{hahm85, rutherford73}. In the Rutherford regime, the FMR model is constitute of the island evolution, torque balance equation, and phase condition~\cite{fitz93}. This model has been used to study magnetic island control using RMPs~\cite{huang15a, huang16a, furukawa}. Recently, an extended FMR model with neo-classical and two-fluid effects~\cite{fitz2019} is used to simulate the RMP-induced ELM suppression experiments in DIII-D~\cite{lux02}. Predictions from those theory models are highly relevant to the interpretation of experimental results on the RMP-induced ELM suppression~\cite{fitz2020, fitz2020c}.


Physically, the linear theory should be recovered from the quasi-linear one when the related effect tends to zero. However, one can find that the linear HKT solution cannot be reduced from the FMR model in the Rutherford regime in all those theories mentioned above~\cite{fitz2019, fitz91a, huang15a, huang16a,  fitz14}. To make a better understanding of FMR, an unified theory uniformly connecting the HKT and Rutherford regimes may be needed. On the other hand, a quasi-linear theory was proposed to reproduce the Rutherford equation and unify the linear exponential and quasi-linear algebraic growth of intrinsic tearing mode~\cite{li1995}. Nonetheless, one of the necessary assumptions used in the derivation, e.g. the expression of the perturbed stream function, may be not appropriate for FMR~\cite{hahm85}. To uniformly connecting the HKT linear and Rutherford quasi-linear solutions of FMR, we should extend the previous quasi-linear approach.

In this work, we propose a new analytical theory to unify the HKT linear solution and the Rutherford quasi-linear regime based on the in-viscid two-field reduced MHD model~\cite{peng15}. Adopting Li's quasi-linear approach~\cite{li1995}, we obtain a closed system of equations for plasma response in the Taylor problem. Transforming the term of quasi-linear perturbed current into the exponential parts of $\psi_1$ (the perturbed flux function) and $\phi_1$ (the perturbed stream function) and using the Laplace transformation, we obtain an integral form of analytical solution for FMR, which uniformly connect the HKT linear and Rutherford quasi-linear solutions. From the solution of FMR, we find that the quasi-linear effect can be described by a single coefficient $K_s\propto S^{8/5} \psi_c^2$, where $S=\frac{\tau_R}{\tau_A}$ and $\psi_c$ are the Lunquist number and amplitude of external magnetic perturbation, respectively. The HKT linear solution for response can be recovered when the index $K_s$ tends to zero. On the other hand, the quasi-linear effect plays a key role on the island growth when $K_s\sim1$. To analyze the FMR solution, we numerically solve the final expression the normalized flux function $\psi_{\rm norm}$. Numerical results show that $\psi_{\rm norm}$ decreases with $K_s$ since the quasi-linear effect is always negative. Our new analytical solution also has been verified by reduced MHD simulation.

Note that we derive the FMR solution in a static plasma in slab geometry. In such an equilibrium, we can make an exactly comparison between the analytical solution and numerical simulation. Our work can be straightforwardly extended to the cylindrical geometry. On the other hand, our work is appropriate in the regime where effects of the plasma flow and viscosity can be neglected. In the rotating plasma, FMR solution obtained in this work should be improved. Even though, our work may provide a necessary foundation to the building of the unified plasma response model including plasma flow and viscosity effects.

The rest of the paper is organized as follows. In Sec. \ref{sec:2}, we introduce the reduced MHD model of the Taylor problem. Sec. \ref{sec:3} is devoted to the analytical derivation and numerical analysis of the FMR solution. Verification of the analytical solution of FMR by reduced MHD simulation is given in Sec. \ref{sec:4}. Finally, we give a summary and discussion in Sec. \ref{sec:5}.

\section{Model of the reduced MHD equations}
\label{sec:2}
In this work, we study the FMR response induced by external magnetic perturbations in a low-$\beta$ plasma based on the two-field reduced MHD model. Using a Cartesian coordinate system $(x, y, z)$ and introducing the flux function $\psi$ and stream function $\phi$, the magnetic field and the velocity can be written as $\vec{B} = B_T\vec{e}_z+\vec{e}_z\times\nabla\psi$ and $\vec{v} = \vec{e}_z\times\nabla\phi$, where $B_T$ is the toroidal magnetic field. The incompressible two-field reduced MHD model governing $\psi$ and $F$ are given, respectively, by

\begin{eqnarray}
&\frac{\partial\psi}{\partial t}+(\vec{e}_z\times\nabla\phi)\cdot\nabla\psi - B_T\partial_z\phi = \eta J_z \label{rmhd1},\\
&\rho(\frac{\partial}{\partial t}+\vec{v}\cdot\nabla)F=\vec{B}\cdot\nabla J_z \label{rmhd2},\\
\end{eqnarray}
where
\begin{eqnarray}
&F = \vec{e}_z\cdot\nabla\times\vec{v} = \nabla_\perp^2\phi = [\frac{\partial^2}{\partial x^2} + \frac{\partial^2}{\partial y^2}]\phi,\\
&J_z = \vec{e}_z\cdot\vec{J} = \frac{1}{\mu_0}\nabla_\perp^2\psi = \frac{1}{\mu_0}[\frac{\partial^2}{\partial x^2} + \frac{\partial^2}{\partial y^2}]\psi,
\end{eqnarray}
and $\rho$ and $\eta$ are the plasma density and resistivity, respectively. In the Taylor problem, the plasma is surrounded by perfect conducting walls at $x = \pm a$ and the boundary perturbation is specified as $x = \pm(a - \delta_{\rm RMP}e^{iky})$, where $\delta_{\rm RMP}$ is the static boundary displacement~\cite{hahm85}. The plasma equilibrium are set as $B_T=0$, $v_{\rm eq}=0$, and $B_{p\rm eq}=B_p\frac{x}{a}$ in the $2-D$ Taylor problem. Then, the equilibrium toroidal current $J_{z\rm eq}=\frac{1}{\mu_0}\frac{B_p}{a}$ is a constant.

\section{Integral solution of the FMR and its numerical analysis}
\label{sec:3}
In this work, we define $\psi = \psi_{\rm eq} + \delta \psi$, $\phi = \delta \phi$, $\delta \psi = \sum_{l=0}^{\infty}\delta \psi_l = \delta \psi_0 + \sum_{l=1}^{\infty}\psi_l \cos(lky)$, and $\delta \phi = \sum_{l=0}^{\infty}\delta \phi_l = \delta \phi_0 + \sum_{l=1}^{\infty}\phi_l \sin(lky)$. Then, Eqs. (\ref{rmhd1}) and (\ref{rmhd2}) can be reduced to the following forms if one neglects the effects of higher harmonic coupling and plasma flow induced by RMPs
\begin{eqnarray}
&\partial_t \delta\psi_1 + \delta\vec{v}_1\cdot\nabla\psi_{\rm eq} + \underbrace {\delta\vec{v}_1\cdot\nabla\delta\psi_0}_{\rm QLMA} = \eta \delta J_{z1} \label{nonlpsi},\\
&\underbrace {\rho\partial_t\delta F_1}_{\rm ITV} = \vec{B}_{\rm eq}\cdot\nabla\delta J_{z1} + \underbrace {\delta\vec{B}_{1}\cdot\nabla J_{z\rm eq}}_{\rm EQJD} + \underbrace {\delta\vec{B}_{0}\cdot\nabla\delta J_{z1}}_{\rm QLMB} + \underbrace {\delta\vec{B}_{1}\cdot\nabla \delta J_{z0}}_{\rm QLJ}\label{nonlphi},\\
&\underbrace {\partial_t \delta\psi_0}_{\rm QLMC} + \left\langle\delta\vec{v}_1\cdot\nabla\delta\psi_1\right\rangle = \eta \delta J_{z0}\label{nonlpsi0},
\end{eqnarray}
where $\left\langle f \right\rangle = \frac{1}{2\pi}\int_0^{2\pi/k}fdy$. For the convenience of discussion, we mark the quasi-linear magnetic and current terms as $\rm QLMA$, $\rm QLMB$, and $\rm QLMC$ and $\rm QLJ$ in Eqs.~(\ref{nonlpsi})-(\ref{nonlpsi0}). To solve the above quasi-linear Eqs.~(\ref{nonlpsi})-(\ref{nonlpsi0}), $\rm QLMC$ should be neglected at first~\cite{li1995}. If one neglects the terms of $\rm ITV$, $\rm QLMA$, and $\rm QLMB$ in Eqs.~(\ref{nonlpsi}) and (\ref{nonlphi}) and adopt the certain assumption of $\phi_1$ (see also in \ref{app2}), the Rutherford equation can be recovered~\cite{rutherford73}. When $\rm ITV$ in Eq.~(\ref{nonlphi}) is included, one can unify the linear exponential and quasi-linear algebraic growth of intrinsic tearing mode. In addition, the terms of $\rm QLMA$ and $\rm QLMB$ should be contained only if $W\Delta' \gg 1$~\cite{li1995}, where $\Delta'=[\frac{1}{\psi_s}\frac{d \psi_1}{dx}]_{x=0}$ is the tearing index and $[f]_{x=0}$ is the jump across the tearing layer around the resonant surface~\cite{furth63a}.



\subsection{Integral form of the analytical FMR solution}
\label{sec:3.1}
In the outer region, we neglect the non-ideal and quasi-linear terms. Then, solution of Eq.~(\ref{nonlphi}) is
\begin{equation}
\psi_1=[\cosh(kx)-\frac{\sinh(kx)}{\tanh(ka)}]\psi_s+[\frac{\sinh(kx)}{\sinh(ka)}]\psi_c \label{solution},
\end{equation}
where $\psi_s= \psi_1(0)$ and $\psi_c  =  \psi_1(x=\pm a) = B_p\delta_{\rm RMP}$. From Eq.~(\ref{solution}), we can rewrite $\Delta'$ as $\psi_s\Delta'=\psi_s\Delta_0'+B_p\delta_{\rm RMP}\Delta_e'=\frac{-2k}{\tanh(ka)}\psi_s+\frac{2k}{\sinh(ka)}\psi_c$.

In the inner region, $\partial_x \gg k$ is assumed. Within appropriate amplitude of the boundary perturbation~\cite{hahm85, wang92a, comisso15b},  the constant-$\psi$ assumption is valid. Note that ${\rm EQJD}=0$ in our equilibrium. Then, Eqs. (\ref{nonlpsi})-(\ref{nonlpsi0}) in the inner region can be simplified to
\begin{eqnarray}
&\partial_t  \psi_1 - kB_p\frac{x}{a} \phi_1 = \eta  J_{z1}\label{nonlinear1},\\
&\rho\partial_t  F_1 = -kB_p\frac{x}{a}  J_{z1} - \frac{k^2}{2\eta } \psi_1^2\partial_x^2 \phi_1\label{nonlinear2}.
\end{eqnarray}
Here, $\rm QLMA$ and $\rm QLMB$ are neglected since ${B}_{\rm peq}=\partial_x\psi_{\rm eq} \gg \delta{B}_{p0}=\partial_x\delta\psi_0$. This approximation will be verified by reduced MHD simulation in the Sec. \ref{sec:sim}.

Since the term of $\rm EQJD$ in Eq.~(\ref{nonlphi}) is essential only when the mode is close to marginality~\cite{militello04a, militello11a}. For similar reasons, neither has been considered in the classical theory of resistive tearing mode~\cite{rutherford73, furth63a}. Then, Eqs. (\ref{nonlinear1}) and (\ref{nonlinear2}) can be straightforwardly extended to the cylindrical geometry.

In $1995$, Li proposed a quasi-linear approach to reproduce the Rutherford equation and unify the linear exponential and quasi-linear algebraic growth of intrinsic tearing mode by assuming $\phi_1 \propto \partial_t\psi_s\Phi(x)$ (see also in \ref{app2})~\cite{li1995}. Nonetheless, such an assumption may be not appropriate for FMR even in the linear limit (see also in \ref{app3}). Different with \cite{li1995}, we do not take assumptions on the expression of $\phi_1$. Instead, Eqs.~(\ref{nonlinear1}) and (\ref{nonlinear2}) in the inner region are solved using the Laplace transform.

Following ~\cite{huang20a}, we define $f_1=\hat f_1 e^{-\varphi_{\rm temp}}$, where $S=\frac{\tau_R}{\tau_A}$, $K = \frac{k^2}{B_p^2}S^{\frac85}$, and $\varphi_{\rm temp}=\int_0^t\frac{k^2}{2\rho\eta}\psi_s^2(t')dt'=\frac12K\int_0^\tau\psi_s^2(\tau')d\tau'$. Then, we transform Eqs.~(\ref{nonlinear1}) and (\ref{nonlinear2}) to the following forms
\begin{eqnarray}
&\partial_t \hat\psi_1  - \frac{k^2}{2\rho\eta }\psi_s^2\hat\psi_1 - kB_p\frac{x}{a}\hat\phi_1 = \eta \hat J_{z1}\label{nonlinear3},\\
&\rho\partial_t \hat F_1 = -kB_p\frac{x}{a}\hat J_{z1}\label{nonlinear4}.
\end{eqnarray}
Introducing the Laplace transform~\cite{hahm85}, we convert Eqs.~(\ref{nonlinear3}) and (\ref{nonlinear4}) to
\begin{eqnarray}
&\frac{\partial^2}{\partial \chi^2}\Psi = \delta_{\rm RI} \Omega(G + \chi U)\label{PsiLp},\\
&\frac{\partial^2}{\partial \chi^2}U - \chi^2 U = \chi G\label{PhiLp},
\end{eqnarray}
where
\begin{eqnarray*}
&\tau_R=\frac{\mu_0a^2}{\eta}, \tau_A=\frac{a}{B_p/\sqrt{\mu_0\rho}}, \delta_{\rm RI}^4 = \frac{s \tau_A^2}{(ka)^2 \tau_R}, \nu = \frac{s}{k^2 \delta_{\rm RI}}, \Omega = \delta_{\rm RI} \tau_R s, \chi = \frac{x}{\delta_{\rm RI} a},\\
&\Psi_s = \frac{k}{B_p}\tilde \psi_s, \Psi_c = \frac{k}{B_p}\tilde \psi_c, G = \frac{k}{B_p}\tilde g, U = - \tilde \phi / \nu,\\
&s\tilde g = s\tilde\psi_s - \mathcal{L}[(\frac{k^2}{2\rho\eta}\psi_s^2)\hat\psi_s], \\
&\tilde \psi = \mathcal{L}[\hat\psi_1] = \displaystyle \int_0^\infty \hat\psi_1 e^{- s t} d t, \\
&\tilde \phi = \mathcal{L}[\hat\phi_1] = \displaystyle \int_0^\infty \hat\phi_1 e^{- s t} d t. \\
\end{eqnarray*}
In \cite{li1995}, Li unify the linear exponential and quasi-linear algebraic growth of tearing mode. Following their quasi-linear approach, we propose an integral form of FMR response which unify HKT linear and Rutherford quasi-linear solutions. Furthermore, Eq.~(\ref{PhiLp}) can be converted to Eq.~(\ref{nonlphiab4}) if we define $U=YG$. Nonetheless, the final expressions of $\phi_1$ are different from our work (Eq.~(\ref{c4})) and previous one (Eq.~(\ref{C1})).


Via the asymptotic matching, we arrive at
\begin{equation}
\frac{3 \Omega}{\sqrt2r_s} G = \Delta_0'\Psi_s + \Delta_e'\Psi_c\label{Gform}.
\end{equation}
To obtain a transparent solution, we rearrange Eq. (\ref{Gform}) to
\begin{equation}
\Psi_s = \frac{\Delta_e'}{-\Delta_0'}\frac{\Psi_c}{1 + \lambda p^\frac54} + \frac12\frac{K}{p}\left\{1 - \frac{1}{1 + \lambda p^\frac54}\right\}\mathcal{L}[\psi_s^2\hat\psi_s]\label{psiform},
\end{equation}
where $p = s\tau_R^{\frac35}\tau_A^{\frac25}$ and $\lambda = \frac{3}{-2ka\Delta_0'}[\frac{1}{4(ka)^2}]^{\frac14}$. Then, the following expression of $\psi_s$ can be obtained by the inverse Laplace transform
\begin{eqnarray}
&\psi_s(\tau) = \frac{\Delta_e'}{-\Delta_0'}e^{-\varphi_{\rm temp}(\tau)}\int_0^\tau G(\tau-\tau')\psi_c(\tau')e^{\varphi_{\rm temp}(\tau')}d\tau' + \frac12Ke^{-\varphi_{\rm temp}(\tau)}\int_0^\tau\psi_s^3(\tau')e^{\varphi_{\rm temp}(\tau')}d\tau'& \nonumber\\
&\qquad - \frac12Ke^{-\varphi_{\rm temp}(\tau)}\int_0^\tau H(\tau-\tau')\psi_s^3(\tau')e^{\varphi_{\rm temp}(\tau')}d\tau'\label{nonpsi},
\end{eqnarray}
where
\begin{eqnarray*}
&G(\tau) = -\frac45[P_Ae^{P_A \tau}  + P_Be^{P_B \tau}] - \frac{\lambda }{\sqrt{2}\pi} \displaystyle \int_0^\infty e^{- u \tau} \frac{u^{\frac54}}{(1 - \sqrt{2} \lambda u^{\frac54} + \lambda^2 u^{\frac52})} d u,\\
&H(\tau) =1 - \frac45[e^{P_A \tau}  + e^{P_B \tau}] + \frac{\lambda }{\sqrt{2}\pi} \displaystyle \int_0^\infty e^{- u \tau} \frac{u^{\frac14}}{(1 - \sqrt{2} \lambda u^{\frac54} + \lambda^2 u^{\frac52})} d u,
\end{eqnarray*}
and $P_{A}=\lambda^{-\frac45}\rm {exp}(+\frac{4 \pi i}{5})$, $P_{B}=\lambda^{-\frac45}\rm {exp}(-\frac{4 \pi i}{5})$, $\tau=t/(\tau_R^{3/5}\tau_A^{2/5})$.

The above integral form of analytical FMR solution, i.e. Eq.~(\ref{nonpsi}), is uniformly valid throughout the entire regimes from the HKT linear to the Rutherford quasi-linear regimes. Note that the linear solution can not be reduced from the modified Rutherford theory with RMP effect~\cite{hahm85} since the term of $\rm ITV$ in Eq.~(\ref{nonlphi}) is neglected in the previous study~\cite{rutherford73}. On the other hand, one cannot expect that the unified theory uniformly connecting HKT linear and Rutherford quasi-linear solution can be obtained directly setting $\psi_s\Delta'=\psi_s\Delta_0'+\psi_c\Delta_e'$ in Eq.~(\ref{lifinal}) since $\phi_1 \propto \partial_t\psi_s\Phi(x)$, which is one of the necessary assumption in \cite{li1995}, may be not appropriate for FMR even in the linear limit.

To take a further step, we define $\psi_{\rm norm}=\frac{-\Delta_0'}{\Delta_e'}\frac{\psi_s}{\psi_c}$, $K_s=\frac12(\frac{\Delta_e'}{-\Delta_0'})^2K\psi_c^2$, $\varphi_s(\tau)=\int_0^\tau\psi_{\rm norm}^2(\tau')d\tau'$, and $\varphi_{\rm temp}=K_s\varphi_s(\tau)$. Then, Eq. (\ref{nonpsi}) can be rearranged to
\begin{eqnarray}
\psi_{\rm norm}(\tau) &= \int_0^\tau G(\tau-\tau')e^{\varphi_K(\tau, \tau')}d\tau' + K_s\int_0^\tau [1-H(\tau-\tau')]\psi_{\rm norm}^3(\tau')e^{\varphi_K(\tau, \tau')}d\tau'& \nonumber\\
\qquad &= \underbrace {\int_0^\tau G(\tau-\tau')d\tau'}_{\rm LIN} + \underbrace {\int_0^\tau G(\tau-\tau')\left\{e^{\varphi_K(\tau, \tau')}-1\right\}d\tau'}_{\rm QLA} + & \nonumber\\
&\qquad \underbrace {K_s\int_0^\tau [1-H(\tau-\tau')]\psi_{\rm norm}^3(\tau')e^{\varphi_K(\tau, \tau')}d\tau'}_{\rm QLB}\label{nonpsinorm},
\end{eqnarray}
where $\varphi_K(\tau, \tau')=K_s[\varphi_s(\tau')-\varphi_s(\tau)]$. To analyze the dynamics of FMR, we divide the expression of $\psi_{\rm norm}$ into three parts, i.e. $\rm LIN$, $\rm QLA$, and $\rm QLB$. Here, $\rm LIN$ is the normalized HKT solution. $\rm QLA$ and $\rm QLB$ represent the quasi-linear effect. Further analysis will be shown in Sec. \ref{sec:num}.

Note that $K_s=\frac12\frac{k^2}{B_p^2}(\frac{\Delta_e'}{-\Delta_0'})^2(\frac{\tau_R}{\tau_A})^{\frac85}\psi_c^2$ can be viewed as the quasi-linear index. When $K_s\rightarrow 0$, the linear HKT solution is recovered~\cite{hahm85}. Then, the normalized HKT linear solution is equivalent to Eq.~(\ref{nonpsinorm}) with $K_s=0$. On the other hand, quasi-linear effect may play an important role if $K_s \sim 1$. Since $K_s \propto S^{8/5}\psi_c^2$, the quasi-linear effect increases with $\psi_c$ and decreases with $\eta$.

Before we analyze the dynamics of FMR solution, it is useful to discuss the steady state solution of Eqs.~(\ref{nonlinear1}) and~(\ref{nonlinear2}). In fact, the steady state of $\phi_1$ satisfies
\begin{equation}
\frac{\partial^2}{\partial x^2}\phi_1-\frac{2B_p^2}{\psi_s^2}\frac{x^2}{a^2}\phi_1=0\label{phiss}.
\end{equation}
A trivial solution of Eq.~(\ref{phiss}) is $\phi_1(x, t\rightarrow \infty)=0$. Then, $J_{z1}(x, t\rightarrow \infty)=0$. Via the asymptotic matching, the steady state expression of $\psi_s$ satisfies $\psi_s = -\frac{\Delta_e'}{\Delta_0'}\psi_c$, or, $\psi_{\rm norm}=1$. Such a steady state solution can also be obtained from the modified Rutherford equation~\cite{hahm85} and verified by numerical evaluation and reduced MHD simulation in the Secs. \ref{sec:num} and \ref{sec:sim}, respectively.

\subsection{Numerical analysis of the FMR solution}
\label{sec:num}
In this subsection, we numerically evaluate Eq. (\ref{nonpsinorm}) and analyze characteristics of FMR. Since the right hand side of Eq.~(\ref{nonpsinorm}) is an integral form of $\psi_{\rm norm}$, we use the Gauss-Seidel iteration to obtain the numerical evaluation of $\psi_{\rm norm}$. The basic parameters used here are $a=0.5m$, $k=1/a$, $\rho=1.67\times10^{-8}kg/m^3$, and $B_p=0.2T$.

\begin{figure}[htbp]
\setlength{\abovecaptionskip}{0.cm}
\setlength{\belowcaptionskip}{-0.cm}
\centering
\begin{center}
\includegraphics[width=10cm]{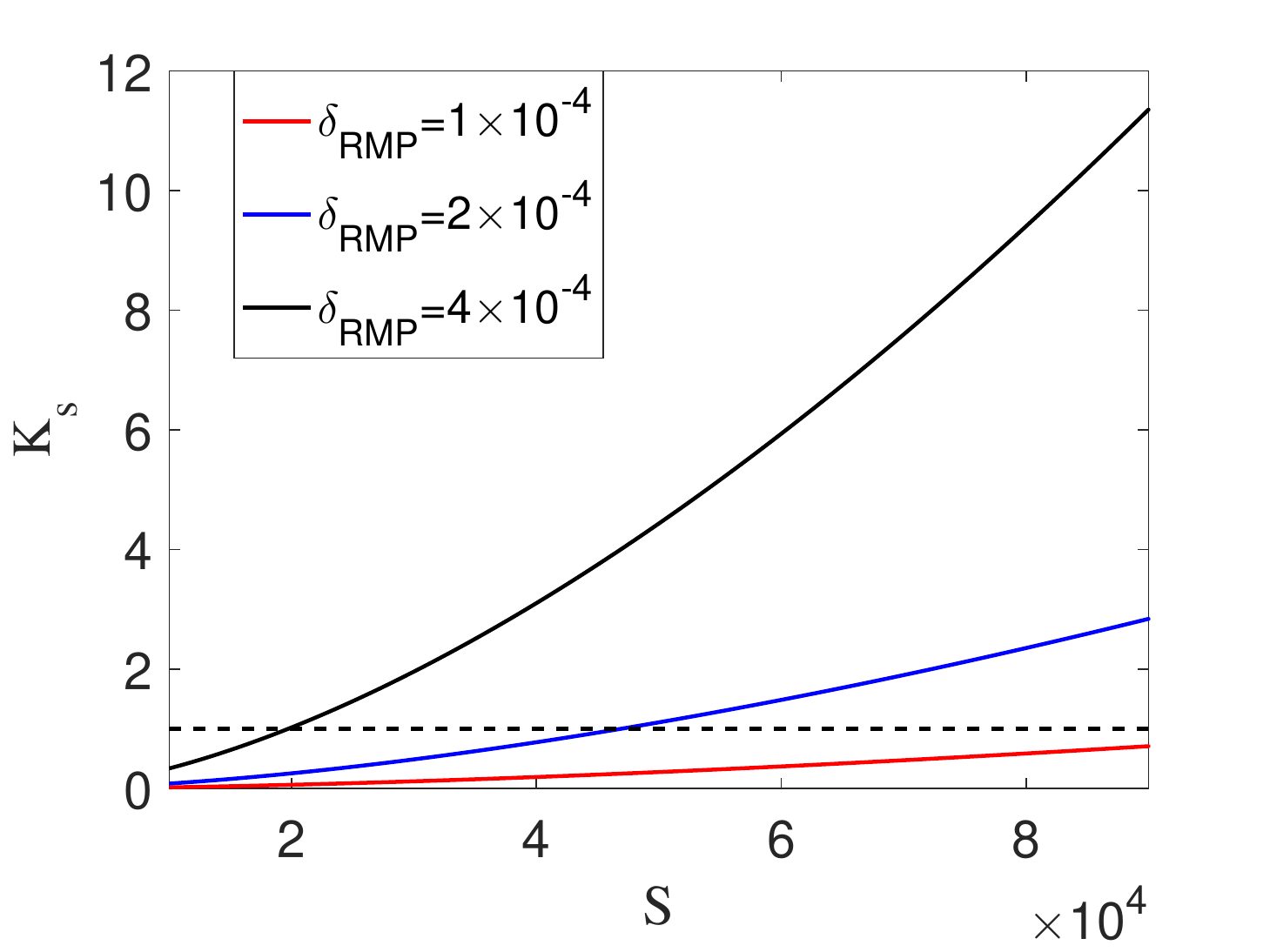}\\
\includegraphics[width=10cm]{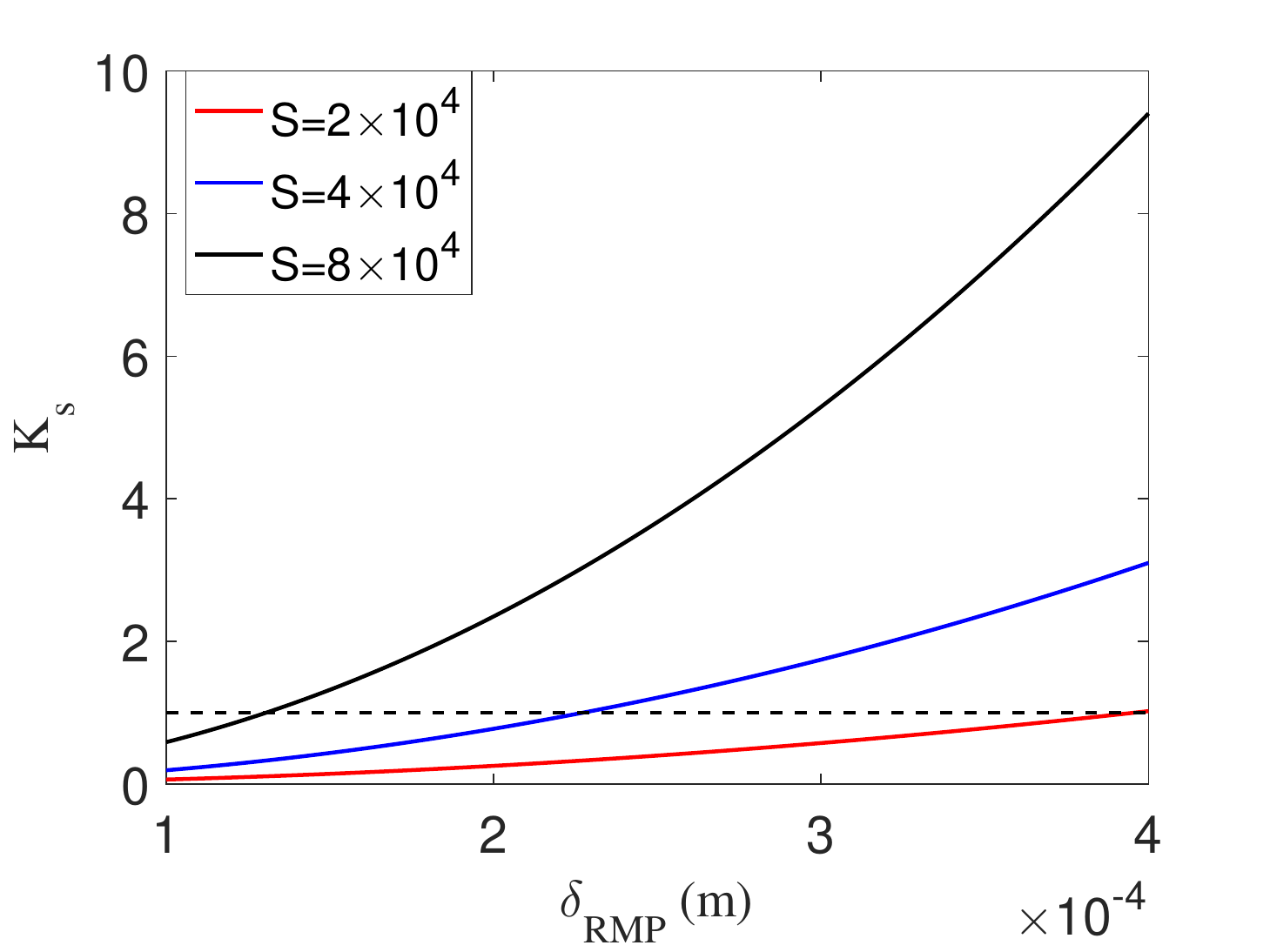}
\end{center}
\caption{The upper and lower panels are dependence of $K_s$ on Lunquist number for different $\delta_{\rm RMP}$ and RMP amplitude for different $S$, respectively. The red, blue, and black curves in the upper (lower) panel represent $\delta_{\rm RMP}=1\times 10^{-4} m$ ($S=2\times 10^{4}$), $\delta_{\rm RMP}=2\times 10^{-4} m$ ($S=4\times 10^{4}$), and $\delta_{\rm RMP}=4\times 10^{-4} m$ ($S=8\times 10^{4}$), respectively. The horizontal dashed lines in the upper and lower panels represent $K_s=1$.}
\label{fig1}
\end{figure}

In Fig.~\ref{fig1}, we illustrate the dependence of $K_s$ on the Lunquist number (RMP amplitude) for different $\delta_{\rm RMP}$ ($S$) to display  parameter regimes where $K_s \ll 1$ and $K_s \sim 1$. For example, $K_s \sim 1$ ($K_s \ll 1$) when $S \sim 2 \times 10^4$ ($S \ll 2 \times 10^4$) with $\delta_{\rm RMP} = 4 \times 10^{-4}m$.

\begin{figure}[htbp]
\setlength{\abovecaptionskip}{0.cm}
\setlength{\belowcaptionskip}{-0.cm}
\centering
\begin{center}
\includegraphics[width=10cm]{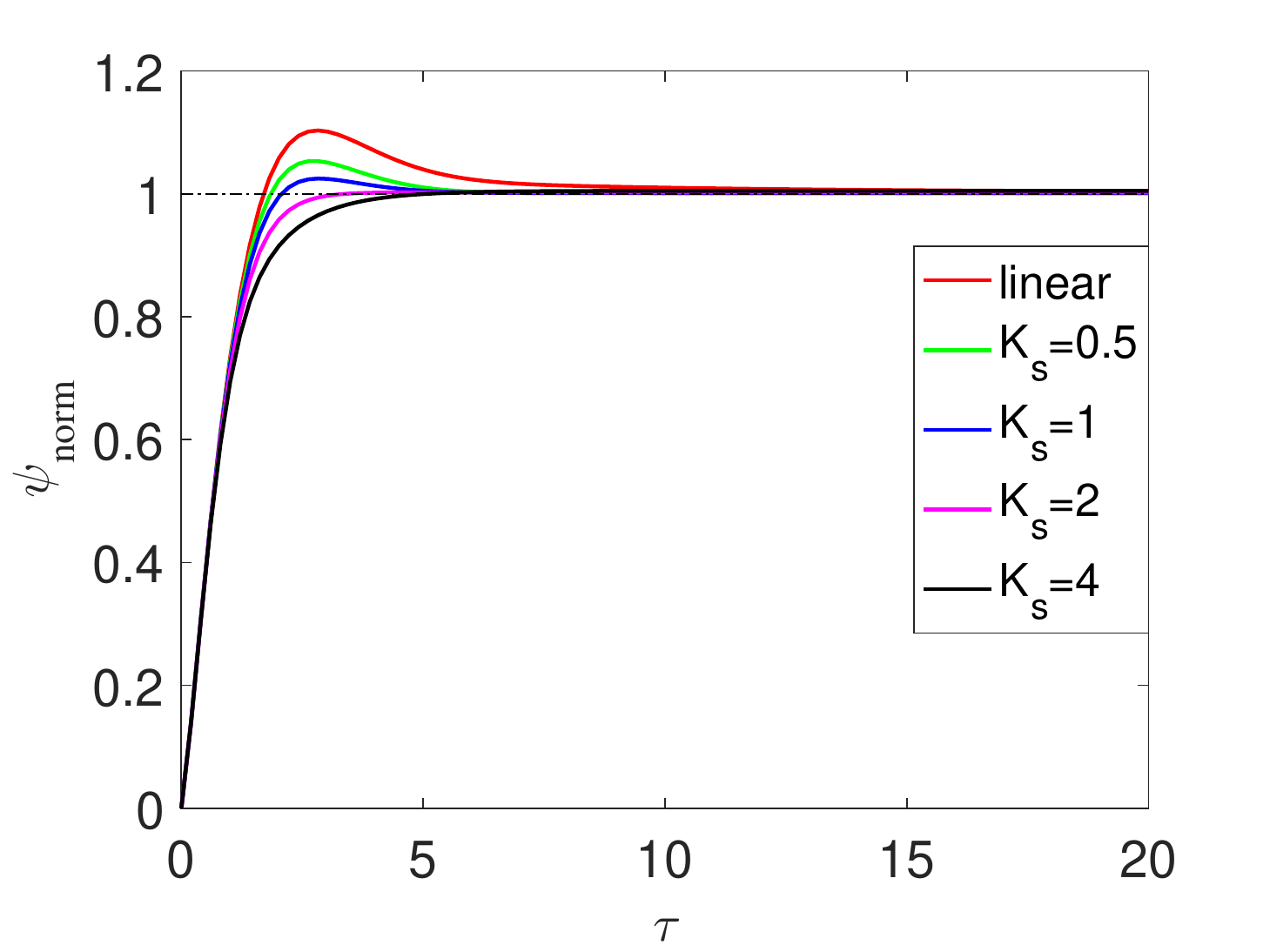}
\end{center}
\caption{The normalized flux function as a function of $\tau$ ($\tau=\frac{t}{\tau_R^\frac35\tau_A^\frac25}$) for different $K_s$ according to Eq. (\ref{nonpsinorm}). The red, green, blue, magenta, and black curves represent $K_s=0$, $K_s=0.5$, $K_s=1$, $K_s=2$, and $K_s=4$, respectively. The black horizontal line is $\psi_{\rm norm} = 1$.}
\label{fig2}
\end{figure}

Time evolution of $\psi_{\rm norm}$ for different $K_s$ is shown in Fig.~\ref{fig2}. When $K_s=0.5$, dynamics of the forced magnetic island is similar with the linear HKT solution. At first, $\psi_{\rm norm}$ rapidly grows with time. When $\tau$ is larger than a critical value, the forced island decreases and tends to a steady state. On the other hand, the overshoot behavior of $\psi_{\rm norm}$ gradually disappears with the increase of $K_s$. Instead, $\psi_{\rm norm}$ increases with $\tau$ and gradually evolves to the steady state. Due to the quasi-linear effect, the forced island width decreases with $K_s$. Besides, the steady state solutions in both the linear and quasi-linear regimes satisfy $\psi_{\rm norm}=1$, which is consistent with our theoretical analysis in the last of Sec.~\ref{sec:3.1}. 

\begin{figure}[htbp]
\begin{center}
\includegraphics[width= 10 cm]{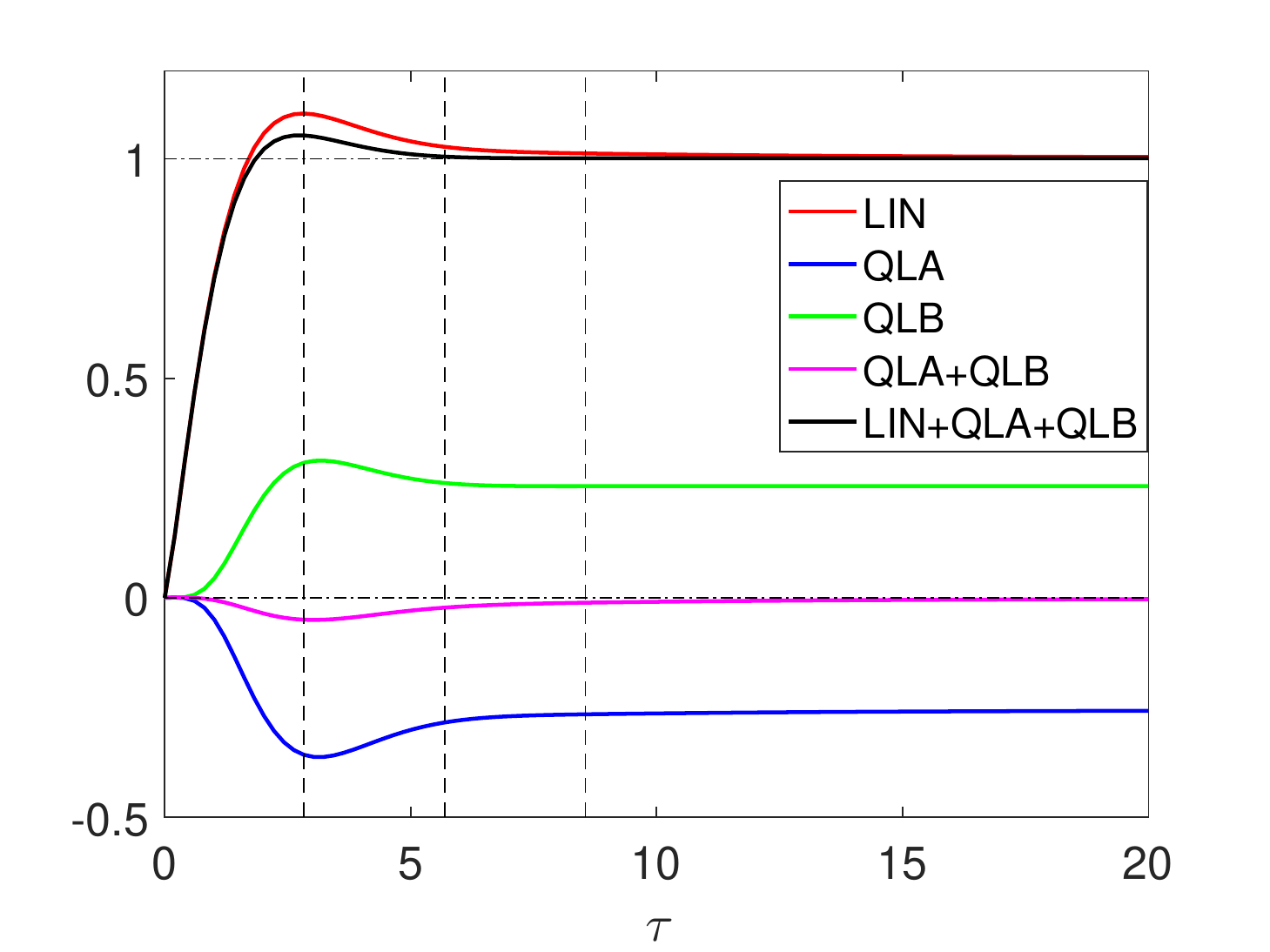}
\includegraphics[width= 10 cm]{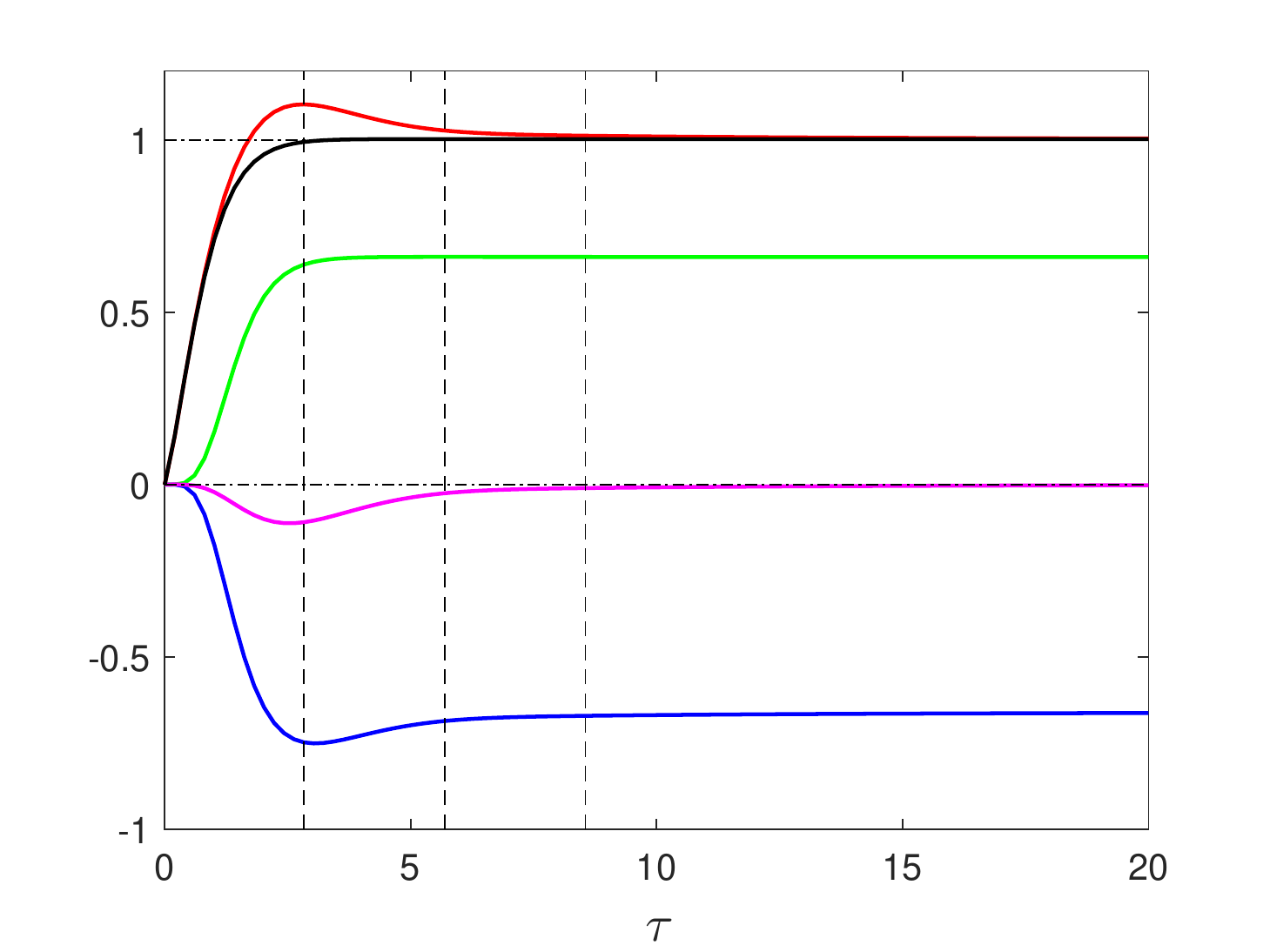}
\caption{The upper and lower panels are different parts of Eq. (\ref{nonpsinorm}) with $K_s=0.5$ and $K_s=2$. The red, blue, green, magenta, and black curves represent $\rm LIN$, $\rm QLA$, $\rm QLB$, $\rm QLA + QLB$, and $\rm LIN + QLA + QLB$, respectively. The black
vertical lines from left to right are $\tau=2.83$, $\tau=5.69$, and $\tau=8.55$, respectively.}
\label{fig3}
\end{center}
\end{figure}

To analyze the quasi-linear effect in further, we plot different parts of Eq.~(\ref{nonpsinorm}) in Fig.~\ref{fig3}. In Fig.~\ref{fig3}, $\rm QLA$ and $\rm QLB$ are negative and positive, respectively. Besides, $|\rm QLA| \geq |\rm QLB|$. Then, the net contribution of quasi-linear terms in Eq.~(\ref{nonpsinorm}) is always negative. Comparisons between the upper and lower panels of Fig.~\ref{fig3} show that $|\rm QLA + \rm QLB|$ increases with $K_s$. These findings explain why $\psi_{\rm norm}$ decreases with $K_s$ in Fig.~\ref{fig2}. On the other hand, $|\rm QLA|$ and $|\rm QLB|$ increase from $0$ to their maximum values in the time scale $\tau \sim 3$. When $\tau \sim 10$, $\rm QLA$ and $\rm QLB$ are constants and canceled by each other. Then, as is shown in Fig.~\ref{fig2}, the steady state of the normalized flux function is $\psi_{\rm norm}=1$. 

\section{Verification of the analytical model using reduced MHD simulation}
\label{sec:4}
In the Sec.~\ref{sec:3}, we propose an unified theory to connect the HKT linear and Rutherford quasi-linear FRM solutions. To verify the integral form of FMR solution in Eq.~(\ref{nonpsinorm}), we develop a two-field reduced MHD code.

\subsection{Details of the two-field reduced MHD simulation code}
In this subsection, we numerically solve the following quasi-linear reduced MHD model

\begin{eqnarray}
&\partial_t \delta\psi_1 + \delta\vec{v}_1\cdot\nabla\psi_{\rm eq} + \delta\vec{v}_1\cdot\nabla\delta\psi_0 = \eta \delta J_{z1} \label{nonlpsia},\\
&\rho\partial_t \delta F_1 = \vec{B}_{\rm eq}\cdot\nabla \delta J_{z1} + \delta\vec{B}_{0}\cdot\nabla \delta J_{z1} + \delta\vec{B}_{1}\cdot\nabla \delta J_{z0}\label{nonlphib},\\
&\eta \delta J_{z0} = \left\langle\delta\vec{v}_1\cdot\nabla\delta\psi_1\right\rangle \label{nonlpsi0c}.
\end{eqnarray}
Here, $\rm QLMC$ is not included in Eq.~(\ref{nonlpsi0c}). The boundary conditions for the stream function and flux function are $\phi_1(\pm a)=0$ and $\psi_1(\pm a) =\psi_c$. Spatial derivatives are calculated by the second order central difference method. The time advance uses the prediction-correction method. The basic numerical parameters used here are $\Delta t = 1.82 \times 10^{-8}s$ and $\Delta x = 0.0025m$. 

\subsection{Simulation results}
\label{sec:sim}
To compare with the analytical form of $\psi_{\rm norm}$, the quasi-linear magnetic terms, i.e. $\rm QLMA$, $\rm QLMB$, and $\rm QLMC$, are neglected at first.

\begin{figure}[htbp]
\begin{center}
\includegraphics[width= 10 cm]{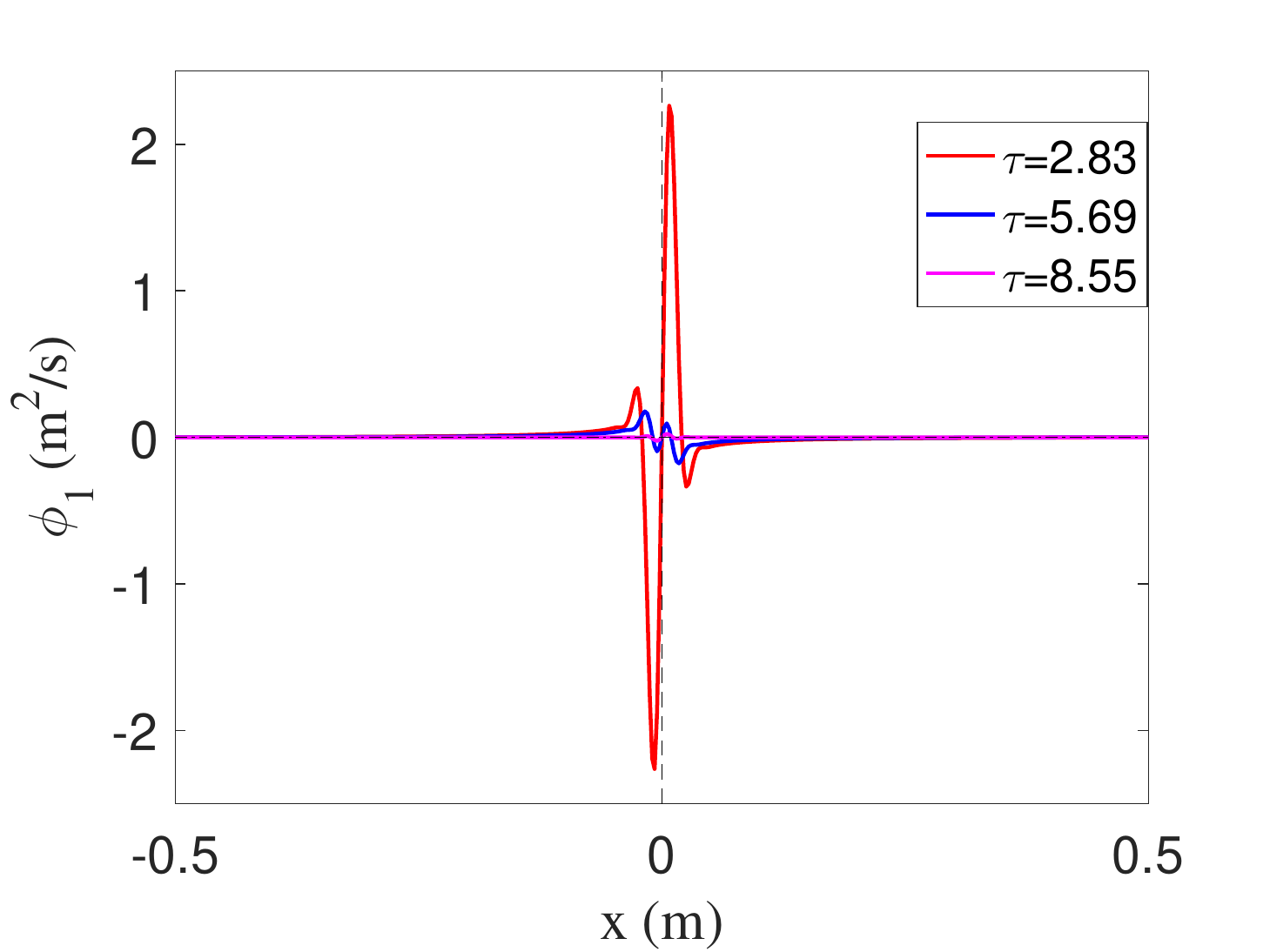}\\
\includegraphics[width= 10 cm]{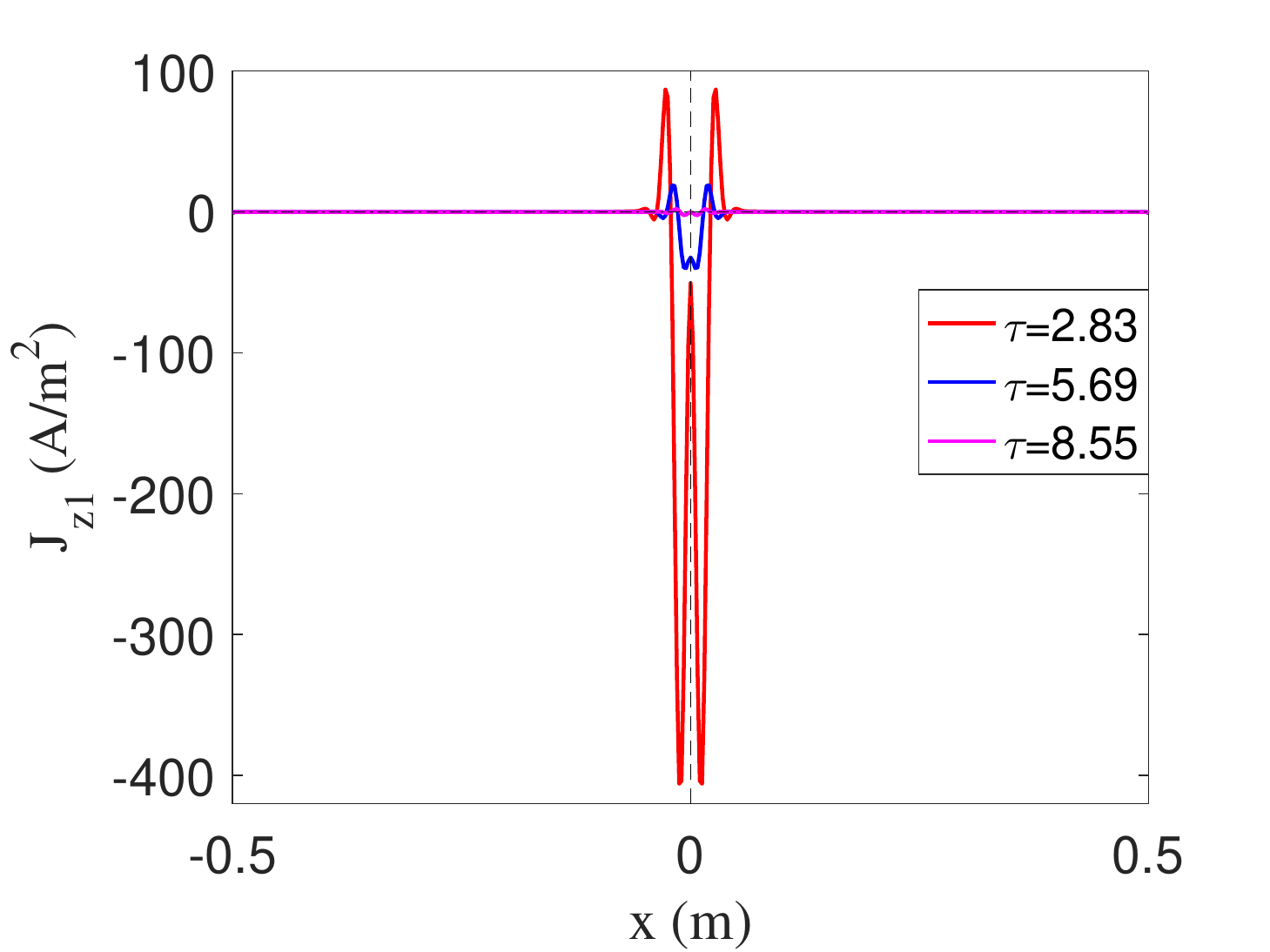}
\caption{The upper and lower panels are distribution of $\phi_1$ and $J_{z1}$ at different time. The red, blue, and magenta curves represent
$\tau=2.83$, $\tau=5.69$, and $\tau=8.55$, respectively.  The plasma resistivity and quasi-linear index used here are $\eta=5\times 10^{-5}\Omega m$ and $K_s=0.5$.}
\label{fig4}
\end{center}
\end{figure}

The radial distributions of $\phi_1$ and $J_{z1}$ at different time are shown in Fig.~\ref{fig4}. In Fig.~\ref{fig4}, $\phi_1$ and $J_{z1}$ are odd and even functions of $x$, respectively. When $\tau \sim 2.83$, $\phi_1$ and $J_{z1}$ reach their maximum values. The perturbed quantities are vanished when $\tau \sim 8.55$. These two properties are consistent with Fig.~\ref{fig3}. In particular, $J_{z1} \rightarrow 0$ when $\tau \sim 8.55$ indicates that the steady state solution of FMR is $\psi_{\rm norm}=1$, as is discussed in Sec. \ref{sec:3}.

\begin{figure}[htbp]
\begin{center}
\includegraphics[width= 7 cm]{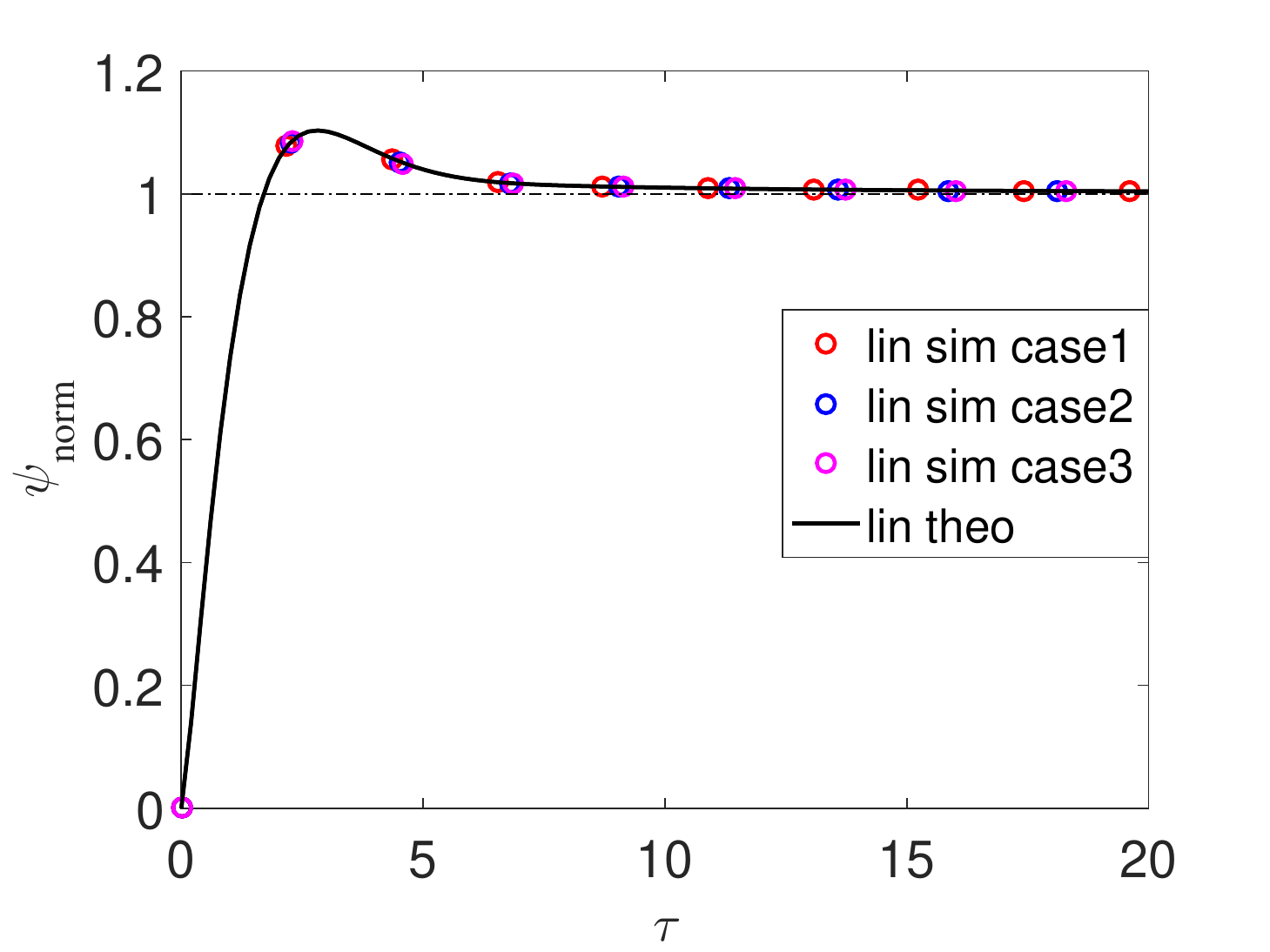}\\
\includegraphics[width= 7 cm]{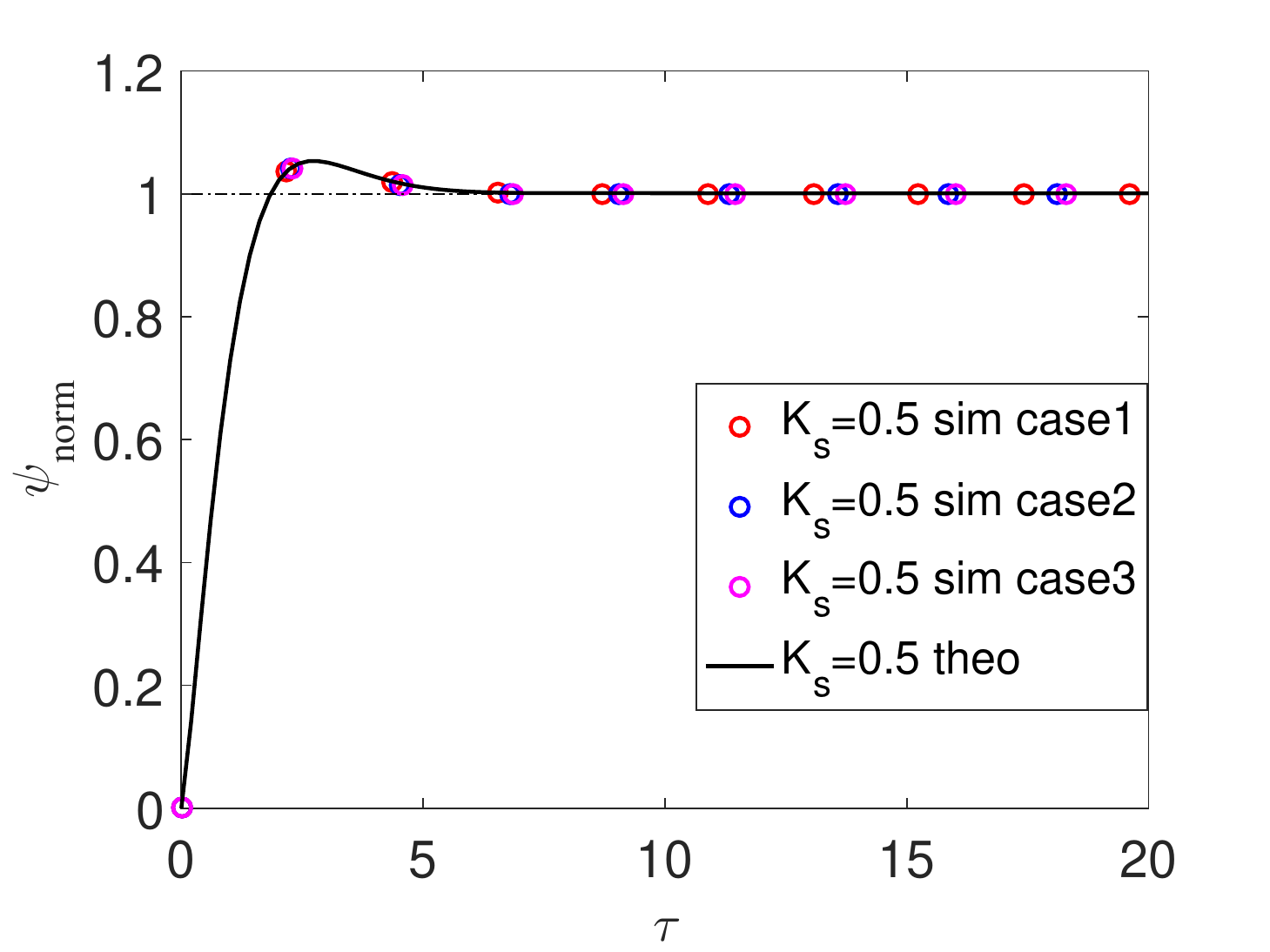}\\
\includegraphics[width= 7 cm]{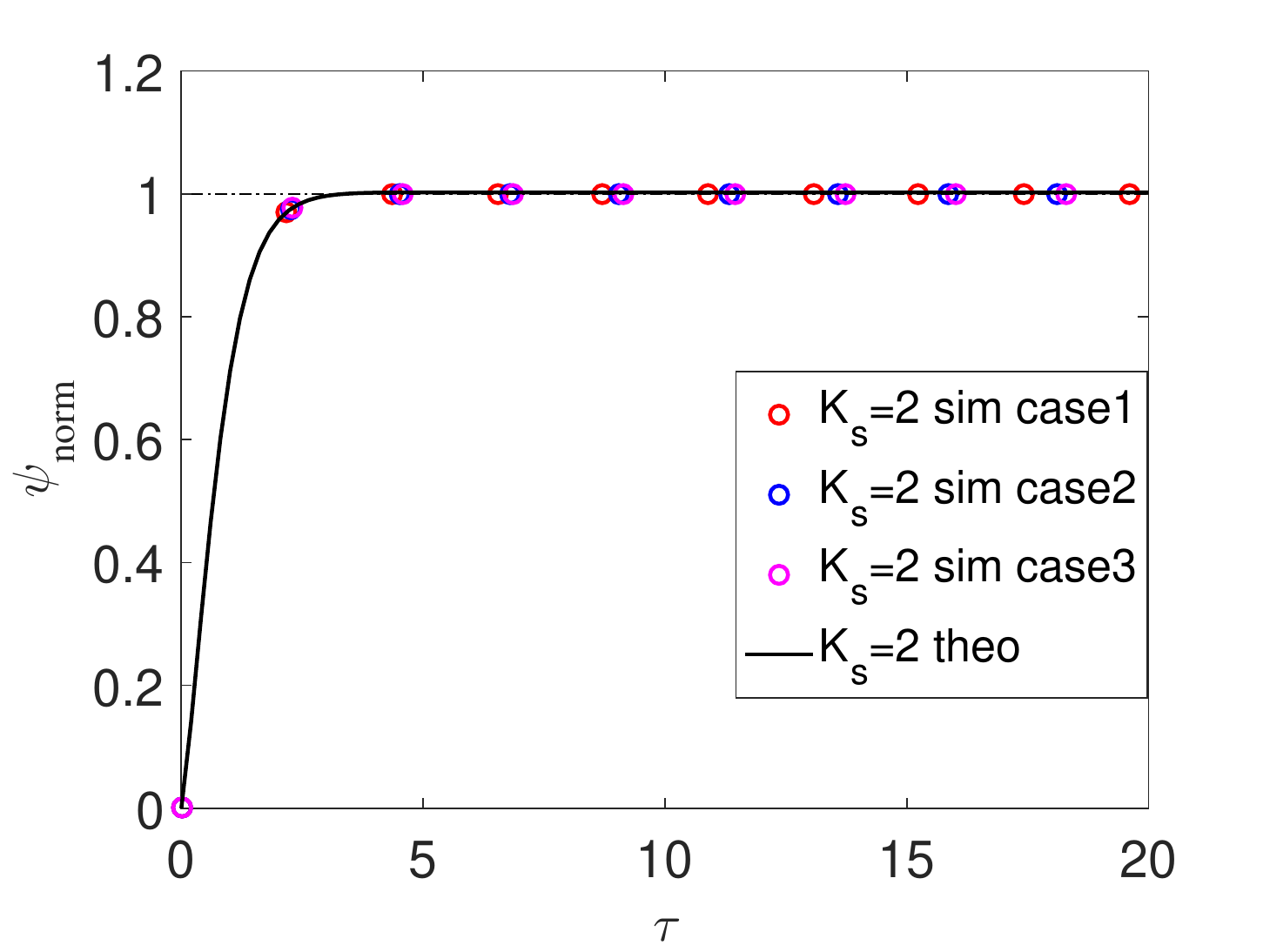}\\
\caption{The top, middle, and bottom panels are dependence of $\psi_{\rm norm}$ on $\tau$ for $K_s=0$, $K_s=0.5$, and $K_s=2$, respectively. The black curves are analytical results calculated from Eq.~(\ref{nonpsinorm}). The red, blue, and magenta cycles are simulation results with $\eta=1\times 10^{-5}\Omega m$ (case $1$), $\eta=2.5\times 10^{-5}\Omega m$ (case $2$), and $\eta=5\times 10^{-5}\Omega m$ (case $3$), respectively.}
\label{fig6}
\end{center}
\end{figure}

Comparison between theory and simulation results of $\psi_{\rm norm}$ for different $K_s$ is shown in Fig.~\ref{fig6}. The theoretical results are calculated from Eq.~(\ref{nonpsinorm}). As is shown in Eq.~(\ref{nonpsinorm}), the quasi-linear effect could be described by a single coefficient $K_s$. To verify the integral form of $\psi_{\rm norm}$, we scan $\eta$ with fixed $K_s$ in simulation. In the top ($K_s=0$), middle ($K_s=0.5$), and bottom ($K_s = 2$) panels of Fig.~\ref{fig6}, simulation results for different $\eta$ are all consistent with theoretical expression of $\psi_{\rm norm}$. This fact indicates that the quasi-linear effect in the reduced MHD simulation can also be described by the quasi-linear coefficient $K_s$.

\begin{figure}[htbp]
\begin{center}
\includegraphics[width= 10 cm]{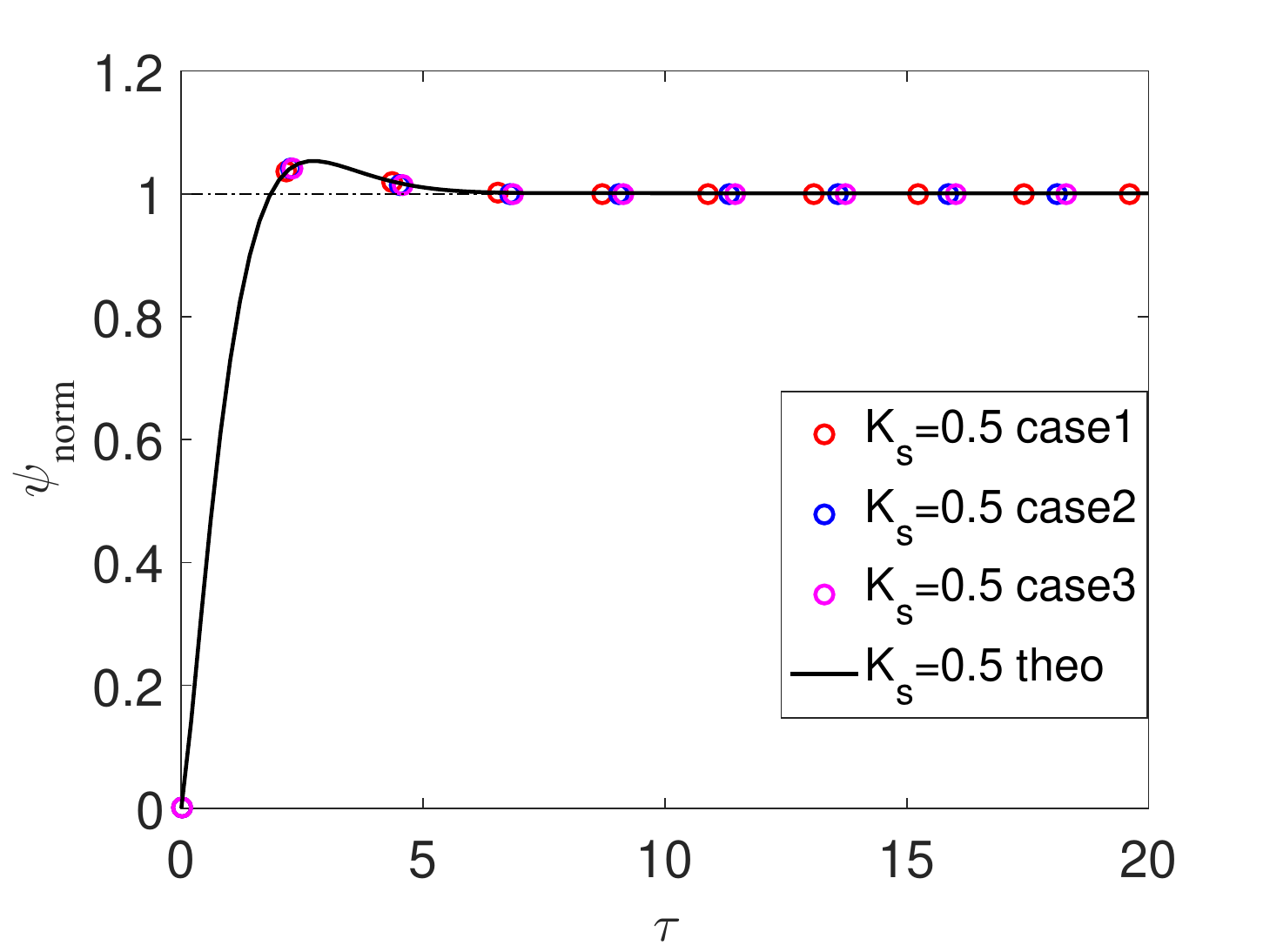}\\
\includegraphics[width= 10 cm]{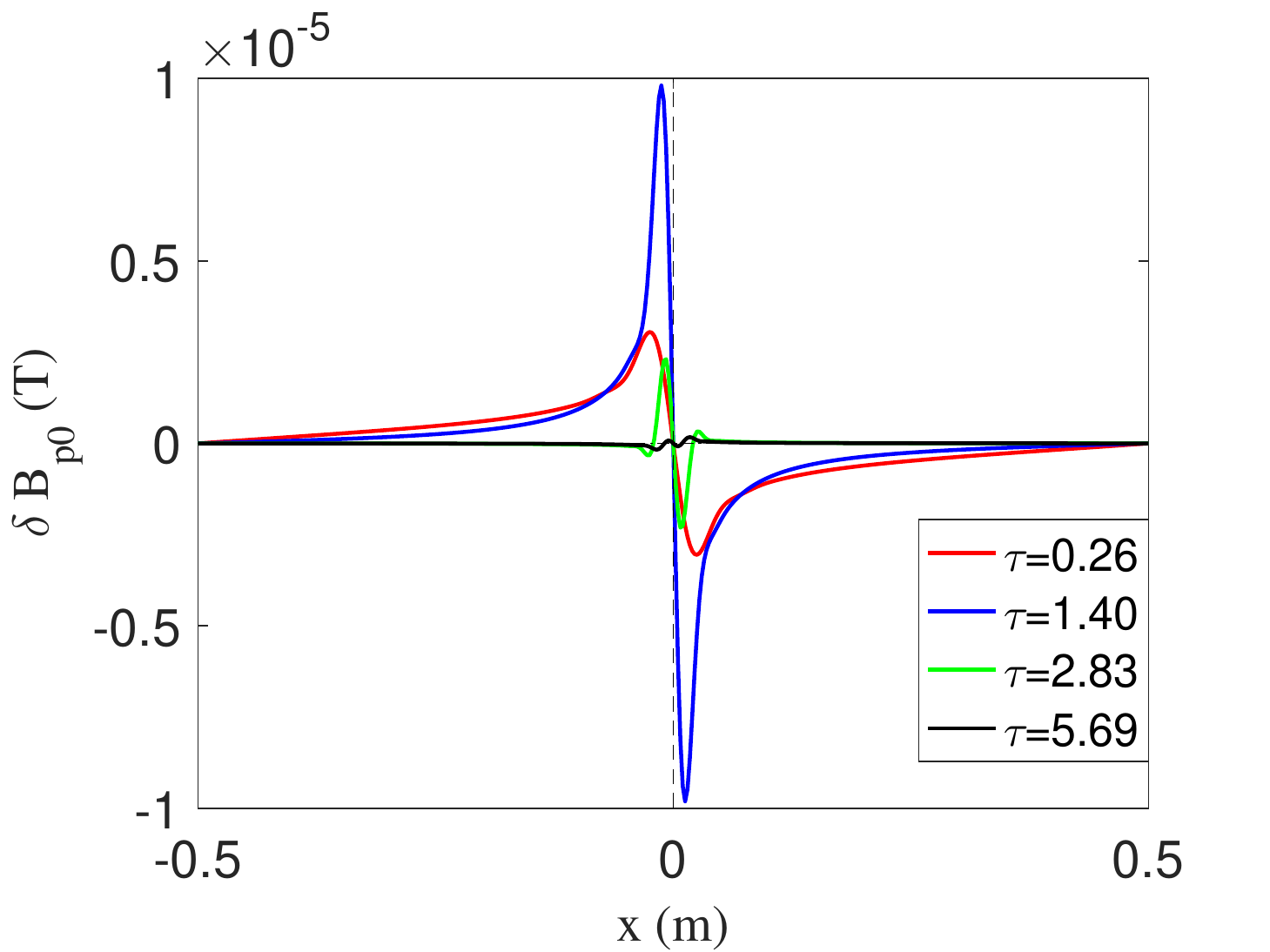}\\
\caption{The upper and lower panels are dependence of $\psi_{\rm norm}$ on $\tau$ with different $\eta$ and distribution of $\delta B_{p0}$ at different time, respectively. The quasi-linear index used here is $K_s=0.5$. The red, blue, and magenta cycles in the upper panel are simulation results of $\psi_{\rm norm}$ for $\eta=1\times 10^{-5}\Omega m$ (case $1$), $\eta=2.5\times 10^{-5}\Omega m$ (case $2$), and $\eta=5\times 10^{-5}\Omega m$ (case $3$), respectively. The black curve is calculated from Eq.~(\ref{nonpsinorm}). In the lower panel ($\eta = 5 \times 10^{-5} \Omega m$), the red, blue, green, and black curves represent $\tau=0.26$, $\tau=1.40$, $\tau=2.83$, and $\tau=5.69$, respectively.}
\label{fig7}
\end{center}
\end{figure}

Next, terms of $\rm QLMA$ and $\rm QLMB$ are added in the simulation code. In the upper panel of Fig.~(\ref{fig7}), we find the simulation results with $\rm QLMA$ and $\rm QLMB$ for different $\eta$ are also agree well with Eq.~(\ref{nonpsinorm}). This can be explained from Eqs.~(\ref{nonlpsia}) and (\ref{nonlphib}). In fact, the related terms in Eqs.~(\ref{nonlpsia}) and (\ref{nonlphib}) can be expressed as $\delta\vec{v}_1\cdot\nabla\psi_{\rm eq} + \delta\vec{v}_1\cdot\nabla\delta\psi_0 = v_{1x}(B_{\rm peq} + \delta B_{p0})$ and $\vec{B}_{\rm eq}\cdot\nabla \delta J_{z1} + \delta\vec{B}_{0}\cdot\nabla \delta J_{z1} = -kJ_{z1}(B_{\rm peq} + \delta B_{p0})$, where $\delta B_{p0} = \partial_x \delta \psi_0$. On the other hand, $B_{\rm peq} \gg \delta B_{p0}$ in the quasi-linear stage. Then, $\rm QLMA$ and $\rm QLMB$ are not important for the island evolution. This conclusion is also verified in the lower panel of Fig.~\ref{fig7}. In this panel, one can find that $\delta B_{p0} \sim 10^{-5} T$. And also, amplitude of the equilibrium magnetic field is $B_{\rm peq} \sim 0.2 T$. Then, $\rm QLMA$ and $\rm QLMB$ can be neglected in the derivation.

It should be noted that $\rm QLMC$ and nonlinear coupling terms are neglected in both theoretical analysis and simulation results. Furthermore, effects of plasma flow and viscosity are not included here. All these effects can may be dramatically influence the dynamics of FMR~\cite{lili} and tearing mode~\cite{zhou} but are well beyond the scope of this work.

\section{Summary and discussion}
\label{sec:5}

In this work, we have proposed a new analytical theory to unify the HKT linear and Rutherford quasi-linear solutions using the in-viscid two-field reduced MHD model. Via the quasi-linear approach~\cite{li1995}, we obtained a closed system of equations for FMR response in the Taylor problem. It should be noted that the quasi-linear approach was proposed to reproduce the Rutherford equation and unify the linear exponential and quasi-linear algebraic growth of intrinsic tearing mode. One of the necessary assumptions used in \cite{li1995}, e.g. the expression of $\phi_1$, may be not appropriate for FMR response. To unify the HKT linear and Rutherford quasi-linear solutions, mathematical skills used in previous work should be extended. Transforming the quasi-linear term, i.e. $\rm QLJ$ in Eq.~(\ref{nonlphi}), into the exponential part of $\phi_1$ and $\psi_1$, the rearranged MHD equations can be solved as before~\cite{huang20a}. Here, the terms of $\rm QLMA$, $\rm QLMB$, and $\rm QLMC$ in Eqs.~(\ref{nonlpsi})-(\ref{nonlpsi0}) are neglected. Note that all those three terms are not needed to reproduce the Rutherford equation. From \cite{li1995}, the terms of $\rm QLMA$ and $\rm QLMB$ are important only in the regime where $W\Delta' \gg 1$. Such a regime is well beyond the scope of this work. In addition, our work can be straightforwardly extended to the cylindrical geometry.

Our FMR solution uniformly valid throughout the entire regimes from the HKT linear solution to the Rutherford quasi-linear solution. In particular, the quasi-linear effect can be described by a single coefficient $K_s$. When the quasi-linear index $K_s\propto S^{8/5} \psi_c^2\rightarrow 0$, the HKT linear solution is recovered. On the other hand, the quasi-linear effect cannot be neglected in the island growth when $K_s\sim1$. We also numerically evaluated the expression of FMR. Results show that the normalized flux function $\psi_{\rm norm}$ decreases with $K_s$ while the steady states of FMR satisfy that $\psi_{\rm norm}=1$. To verify the analytical solution of FMR, we developed a reduced MHD code. Results in our new analytical solution show good agreement with reduced MHD simulations. For example, when $K_s$ is fixed in the reduced MHD code, we found that simulation results for different $\eta$ are all agree with the theoretical expression of $\psi_{\rm norm}$. Then, the quasi-linear effect in the reduced MHD simulation can also be described by the quasi-linear coefficient $K_s$. In addition, the effect of $\rm QLMA$ and $\rm QLMB$ on the FMR response can be neglected, which is consistent with our theoretical analysis.

Due to the limitation of the two-field reduced MHD model, many physics elements for the RMP-induced plasma response have not been included. For example, two-fluid and neo-classical effects are known to have strong influence over plasma response to RMPs near resonant surfaces\cite{fitz2019, fitz2020, fitz2020c, wael12}. Furthermore, our derivation is appropriate only in the static plasma. We plan to address these important issues in future studies.

\ack

We thank Professor Ding Li for his helpful discussion. This work was supported by the Fundamental Research Funds for the Central Universities at Huazhong University of Science and Technology Grant No. 2019kfyXJJS193, the National Natural Science Foundation of China Grant Nos. 11775221 and No. 51821005, the Young Elite Scientists Sponsorship Program by CAST Grant No. 2017QNRC001, and U.S. Department of Energy Grant Nos. DE-FG02-86ER53218 and DE-SC0018001.

\newpage
\begin{appendix}
\section{Derivation procedure of the Rutherford equation}
\label{app1}
To study the island evolution in the regime where $W \sim \delta_{\rm layer}$, a pioneer theory was proposed by Rutherford~\cite{rutherford73}. In this appendix, we give a short review of the derivation of Rutherford equation. Using a Cartesian coordinate system $(x, y, z)$ and introducing the flux function $\psi$ and stream function $\phi$, the magnetic field and the velocity can be written as $\vec{B} = \vec{e}_z\times\nabla\psi$ and $\vec{v} = \vec{e}_z\times\nabla\phi$. The incompressible two-field reduced MHD model governing $\psi$ and $F$ are given, respectively, by
\begin{eqnarray}
&\partial_t \psi + \vec{v}\cdot\nabla\psi= \eta J_{z} \label{nonlpsiap},\\
&\rho(\partial_t + \vec{v}\cdot\nabla)F = \vec{B}\cdot\nabla J_{z}\label{nonlphiap}.
\end{eqnarray}
In the Rutherford regime, the inertial term in Eq.~(\ref{nonlphiap}) can be neglected~\cite{rutherford73}. This assumption leads to $\vec{B}\cdot\nabla J_{z}=0$, implying
\begin{equation}
J_z=J_z(\psi).
\end{equation}
As is shown in \cite{li1995}, see also in \ref{app2}, island evolution in such a regime cannot reduce to the linear exponential one since the inertial term is neglected. To unify the linear exponential and quasi-linear algebraic solution of tearing mode, one should contain the inertial term in  Eq.~(\ref{nonlphiap}). We also define $\psi =\psi (x, y, t)= \psi_{\rm eq}(x) + \delta \psi(x, y, t) = \psi_{\rm eq}(x) + \delta \psi_0 + \sum_{l=1}^{\infty} \psi_l\cos(lky)$ and $\phi =\phi (x, y, t)= \delta \phi(x, y, t) = \delta \phi_0 + \sum_{l=1}^{\infty} \phi_l\sin(lky)$ as usual. In the inner region, $B_{y\rm eq} = \frac{d\psi_{\rm eq}}{dx} \approx B_y'x$. Then, Eq.~(\ref{nonlphiap}) reduces to
\begin{eqnarray}
\partial_t \delta\psi - \frac{\partial \delta\phi}{\partial y}B_y'x= \eta (J_{z}-J_{z\rm eq}) \label{nonlpsiap2}.
\end{eqnarray}
To eliminate $\delta\phi$, we divide $x$ and average over $y$ at constant $\psi$ on Eq.~(\ref{nonlpsiap2})~\cite{rutherford73}. Then, the following equation can be obtained
\begin{eqnarray}
J_z(\psi)- J_{z\rm eq} = \eta^{-1}\left\langle \frac{\partial_t \delta\psi}{[\psi-\delta\psi]^\frac12}\right\rangle/\left\langle[\psi-\delta\psi]^{-1/2}\right\rangle \label{nonlpsiap3},
\end{eqnarray}
where $\left\langle f \right\rangle = \frac{1}{2\pi}\int_0^{2\pi/k}fdy$. 

Via the asymptotic matching, we obtain
\begin{equation}
\Delta_l'\psi_{l} = 2\mu_0\left\langle\cos(lky)\int_{-\infty}^\infty\delta J_zdx\right\rangle\label{nonlpsiap4},
\end{equation}
where $\Delta_l' = [\partial\ln\psi_l/\partial x]_{0-}^{0+}$. Combing Eqs.~(\ref{nonlpsiap3}) and (\ref{nonlpsiap4}), the Rutherford equation can be obtained as following
\begin{equation}
\Delta'\psi_s = \frac{4A\mu_0}{\eta(2B_{y}')^\frac12}\psi_s^\frac12\frac{\partial\psi_s}{\partial t}\label{nonlpsiap5},
\end{equation}
where $\psi_s=\psi_1(x=0)$, $\Delta'=\Delta_1'$, and $A=A_R \approx 0.7$.

\section{Unifying the linear exponential and Rutherford algebraic growth of tearing mode using Li's method}
\label{app2}
To connect the linear exponential and quasi-linear algebraic growth of tearing mode, Li proposed an unique quasi-linear approach~\cite{li1995}. In this appendix, we give a short review of Li's approach. Note that Li developed his approach in cylindrical geometry. To make a comparison with this work, we extend the original method to the two-dimensional slab geometry. We define $\psi = \psi_{\rm eq} + \delta \psi$, $\phi = \delta \phi$, $\delta \psi = \sum_{l=0}^{\infty}\delta \psi_l = \delta \psi_0 + \sum_{l=1}^{\infty}\psi_l \cos(lky)$, and $\delta \phi = \sum_{l=0}^{\infty}\delta \phi_l = \delta \phi_0 + \sum_{l=1}^{\infty}\phi_l \sin(lky)$. Neglecting the higher harmonic coupling, Eqs. (\ref{rmhd1}) and (\ref{rmhd2}) can be reduced to the following forms
\begin{eqnarray}
&\partial_t \delta\psi_1 + \delta\vec{v}_1\cdot\nabla\psi_{\rm eq} + \underbrace {\delta\vec{v}_1\cdot\nabla\delta\psi_0}_{\rm QLMA} = \eta \delta J_{z1} \label{nonlpsiab1},\\
&\underbrace {\rho\partial_t \delta F_1}_{\rm ITV} = \vec{B}_{\rm eq}\cdot\nabla \delta J_{z1} + \underbrace {\delta\vec{B}_{1}\cdot\nabla J_{z\rm eq}}_{\rm EQJD} + \underbrace {\delta\vec{B}_{0}\cdot\nabla \delta J_{z1}}_{\rm QLMB} + \underbrace {\delta\vec{B}_{1}\cdot\nabla \delta J_{z0}}_{\rm QLJ}\label{nonlphiab1},\\
&\underbrace {\partial_t \delta\psi_0}_{\rm QLMC} + \left\langle\delta\vec{v}_1\cdot\nabla\delta\psi_1\right\rangle = \eta \delta J_{z0}\label{nonlpsi0ab1}.
\end{eqnarray}
Usually, effect of $\rm EQJD$ can be neglected in the inner region~\cite{militello04a, militello11a}.

To obtain the Rutherford equation, one should neglect terms of $\rm QLMA$, $\rm QLMB$, $\rm QLMC$, and $\rm ITV$ in Eqs.~(\ref{nonlpsiab1})-(\ref{nonlpsi0ab1}) in the inner region~\cite{li1995}. Then, one arrives at
\begin{eqnarray}
&\partial_t \delta\psi_1 + \delta\vec{v}_1\cdot\nabla\psi_{\rm eq} = \eta \delta J_{z1} \label{nonlpsiab2},\\
&0 = \vec{B}_{\rm eq}\cdot\nabla \delta J_{z1} + \frac{1}{\eta}\delta\vec{B}_{1}\cdot\nabla \left\langle\delta\vec{v}_1\cdot\nabla\delta\psi_1\right\rangle\label{nonlphiab2}.
\end{eqnarray}
Within the constant-$\psi$ assumption, Eqs.~(\ref{nonlpsiab2}) and (\ref{nonlphiab2}) can be rearranged to
\begin{eqnarray}
&\partial_t \psi_s - kB_y'x\phi_1 = \eta J_{z1} \label{nonlpsiab3},\\
&kB_y'x J_{z1} + \frac{k^2}{2\eta}\psi_s^2\partial_x^2\phi_1=0\label{nonlphiab3},
\end{eqnarray}
where $\psi_s=\psi_1(x=0)$. Combing Eqs.~(\ref{nonlpsiab3}) and (\ref{nonlphiab3}) and setting $\phi_1=-\nu\partial_t\psi_sY$, $Y=Y(\chi)$, and $x=\delta\chi$, one can obtain
\begin{eqnarray}
&\frac{d^2Y}{d\chi^2}-\chi^2Y=\chi\label{nonlphiab4},\\
&\delta^4=\frac{\frac12\psi_s^2}{(B_y')^2},\\
&\nu=\frac{1}{kB_y'\delta}.
\end{eqnarray}
Solution of Eq.~(\ref{nonlphiab4}) is~\cite{li1995}
\begin{equation}
Y=-\frac{\chi}{2}\int_0^1d\lambda(1-\lambda^2)^{-\frac14}\exp(-\frac{\lambda\chi^2}{2})\label{nonlphiab5}.
\end{equation}
After asymptotic matching, the final expression of $\psi_s$ is obtained
\begin{eqnarray}
\Delta'\psi_s&=\mu_0\int J_{z1}dx \nonumber\\
\qquad &=-\frac{\mu_0}{2\eta}\frac{k^2}{kB_y'}\psi_s^2\int\frac{\partial_x^2\phi_1}{x}dx \nonumber\\
\qquad &=\frac{C_0\mu_0}{\eta}\delta\partial_t\psi_s \nonumber\\
\qquad &=\frac{4A\mu_0}{\eta}\frac{\psi_s^\frac12}{\sqrt{2B_y'}}\partial_t\psi_s,\\
\end{eqnarray}
where
\begin{eqnarray}
&C_0=\int_{-\infty}^\infty[1+\chi Y]d\chi,\\
&A=A_L=\frac{2^\frac14}{4}C_0\approx\frac{2^\frac14}{4} \times 2.12\approx0.63.
\end{eqnarray}
Obviously, the Rutherford equation can be recovered via this quasi-linear approach except that $A=A_R\approx 0.7$ while $A=A_L\approx0.63$ in Rutherford's and Li's approach, respectively. In addition, Li found a new algebraic growth of tearing mode when $\rm QLMA$ and $\rm QLMB$ are considered.

To connect the linear exponential and quasi-linear algebraic growth of tearing mode, $\rm ITV$ in Eq.~(\ref{nonlphiab1}) should be included. Then, one arrives at
\begin{eqnarray}
&\partial_t \delta\psi_1 + \delta\vec{v}_1\cdot\nabla\psi_{\rm eq} = \eta \delta J_{z1} \label{nonlpsiab6},\\
&\rho\partial_t \delta F_1 = \vec{B}_{\rm eq}\cdot\nabla \delta J_{z1} + \underbrace {\frac{1}{\eta}\delta\vec{B}_{1}\cdot\nabla \left\langle\delta\vec{v}_1\cdot\nabla\delta\psi_1\right\rangle}_{\rm QLJ}\label{nonlphiab6}.
\end{eqnarray}
To take a further step, we simplify Eqs.~(\ref{nonlpsiab6}) and (\ref{nonlphiab6}) to
\begin{eqnarray}
&\partial_t  \psi_s - kB_y'x \phi_1 = \eta  J_{z1}\label{nonlinear7},\\
&\rho\partial_t  F_1 = -kB_y'x J_{z1} - \frac{k^2}{2\eta } \psi_s^2\partial_x^2 \phi_1\label{nonlinear7}.
\end{eqnarray}
Setting $\phi_1=-\frac{1}{kB_y'}\partial_t\psi_s\Phi$ and $\Phi=\Phi(x)$, one obtains
\begin{equation}
\frac{d^2\Phi}{dx^2}[\partial_t+\frac{k^2}{2\rho\eta}\psi_s^2](\partial_t\psi_s)-\frac{(kB_y')^2}{\rho\eta}x^2\partial_t\psi_s\Phi=\frac{(kB_y')^2}{\rho\eta}x\partial_t\psi_s\label{nonlinear8}.
\end{equation}
Introducing $x=\delta_{\rm Li}\chi$ and $\Phi=\delta_{\rm Li}^{-1}Y(\chi)$, then Eq.~(\ref{nonlinear8}) can be simplified to Eq.~(\ref{nonlphiab4}) and
\begin{equation}
\delta_{\rm Li}^4=\frac{\rho\eta}{(kB_y')^2\partial_t\psi_s}[\partial_t+\frac{k^2}{2\rho\eta}\psi_s^2](\partial_t\psi_s).
\end{equation}
Via asymptotic matching, the following expression of $\psi_s$ is obtained
\begin{eqnarray}
\Delta'\psi_s=\mu_0\int J_{z1}dx=\frac{C_0\mu_0}{\eta}\delta_{\rm Li}\partial_t\psi_s,
\end{eqnarray}
which leads to
\begin{eqnarray}
(\Delta')^4=\frac{\rho\eta}{(kB_y')^2}[\partial_t\ln\psi_s+\frac{k^2}{2\rho\eta}\psi_s^2](\frac{C_0\mu_0}{\eta}\partial_t\ln\psi_s)^4\label{lifinal}.
\end{eqnarray}
The linear and Rutherford regimes can be recovered when $\partial_t\ln\psi_s \gg \frac{k^2}{2\rho\eta}\psi_s^2$ and $\partial_t\ln\psi_s \ll \frac{k^2}{2\rho\eta}\psi_s^2$, respectively.



\section{Comparison with a previous theory}
\label{app3}
In~\cite{li1995}, Li unify the linear exponential and quasi-linear algebraic growth of tearing mode. Following their quasi-linear approach, we propose an integral form of FMR response which unify HKT linear and Rutherford quasi-linear solutions. Furthermore, one can find that Eq.~(\ref{PhiLp}) and Eq.~(\ref{nonlphiab4}) share the same form if we define $U=YG$. Then, it is necessary to clarify the difference between our work and previous one.

To solve the reduced MHD equations in the inner region, Li assumes that $\phi_1=-\frac{1}{kB_y'}\partial_t\psi_s\Phi$ and $\Phi=\frac{Y}{\delta_{\rm Li}}$, where $\chi=\frac{x}{\delta_{\rm Li}}$, $\Phi=\Phi(x)$, and $Y=Y(\chi)$~\cite{li1995}. Then,
\begin{eqnarray}
&\phi_1 = -\frac{1}{kB_y'}\partial_t\psi_s\delta_{\rm Li}^{-1}Y(\chi)=-\frac{1}{kB_y'}\partial_t\psi_s\delta_{\rm Li}^{-1}Y(\chi)\label{C1},\\
&\delta_{\rm Li}^4=\frac{\rho\eta}{(kB_y')^2\partial_t\psi_s}[\partial_t+\frac{k^2}{2\rho\eta}\psi_s^2](\partial_t\psi_s).
\end{eqnarray}
When $W \ll \delta_{\rm Li}$, $\delta_{\rm Li}$ satisfies that
\begin{eqnarray}
\delta_{\rm Li}^4=\frac{\rho\eta}{(kB_y')^2\partial_t\psi_s}\partial_t^2\psi_s=\frac{a^4\tau_A^2}{(ka)^2\tau_R}\frac{\partial_t^2\psi_s}{\partial_t\psi_s}\approx\frac{a^4\tau_A^2}{(ka)^2\tau_R}\frac{\partial_t\psi_s}{\psi_s}.
\end{eqnarray}

For the FMR response, equations in the inner region should be solved using the Laplace transform~\cite{hahm85}. From our derivation, $\phi_1$ satisfies that
\begin{eqnarray}
\phi_1 =-\frac{1}{kB_p}\mathcal{L}^{-1}[\frac{s\tilde g}{\delta_{\rm RI}} Y(\chi)]=-\frac{1}{kB_p}\mathcal{L}^{-1}[\frac{s\tilde\psi_s - \mathcal{L}[(\frac{k^2}{2\rho\eta}\psi_s^2)\hat\psi_s]}{\delta_{\rm RI}} Y(\chi)]\label{c4}.
\end{eqnarray}
In the linear limit, $\phi_1$ can be expressed as
\begin{eqnarray}
\phi_1 = -\frac{1}{kB_p}\mathcal{L}^{-1}[\frac{s\mathcal{L}(\psi_s)}{\delta_{\rm RI}} Y(\chi)]=-\frac{1}{kaB_y'}\mathcal{L}^{-1}[\frac{s\mathcal{L}(\psi_s)}{\delta_{\rm RI}} Y(\frac{x}{a\delta_{\rm RI}})]\label{C4a},
\end{eqnarray}
where $B_y'=\frac{B_p}{a}$. If $\delta_{\rm RI}$ does not dependent on $s$, i.e. the Laplace factor, one can rewrite Eq.~(\ref{C4a}) as
\begin{eqnarray}
\phi_1 =-\frac{1}{kB_y'}\frac{Y(\chi)}{a\delta_{\rm RI}}\mathcal{L}^{-1}[s\mathcal{L}(\psi_s)]\label{C4}=-\frac{1}{kB_y'}\frac{Y(\chi)}{a\delta_{\rm RI}}\frac{\partial\psi_s}{\partial t}\label{C5b}.
\end{eqnarray}
It is worth noting that $\delta_{\rm RI}^4=\frac{s\tau_A^2}{(ka)^2\tau_R}$, i.e. $\delta_{\rm RI} \propto s^\frac14$, or, $\delta_{\rm RI}$ is clearly dependent on $s$. Then, one cannot reduce Eq.~(\ref{C4a}) to Eq.~(\ref{C5b}). The final solutions of $\phi_1$ are different from our work and previous one. Then, one of the necessary assumptions used in \cite{li1995}, i.e. expression of $\phi_1$, may be not appropriate for FMR even in the linear limit.

Based on above the discussion, one cannot expect that the unified theory uniformly connecting HKT linear and Rutherford quasi-linear solution can be obtained directly setting $\psi_s\Delta'=\psi_s\Delta_0'+\psi_c\Delta_e'$ in Eq.~(\ref{lifinal}).

\end{appendix}

\newpage
\section*{References}
\providecommand{\newblock}{}

\newpage
\newpage
\newpage

\newpage

%
\newpage

\newpage

%

%

\end{document}